\documentclass{moriond}
\usepackage{amsmath,amssymb,epsfig,ulem,color,slashed,multirow,wasysym}

\def\be{\begin{equation}}
\def\ee{\end{equation}}
\def\bea{\begin{eqnarray}}
\def\eea{\end{eqnarray}}

\def \bm#1{\mbox{\boldmath$#1$\unboldmath}} 

\newcommand{\Photo}{}

\newcommand{\MET}{E_{T,{\rm miss}}}

\begin{document}
\title{Dark matter at the LHC}

\author{Ulrich Haisch }

\address{Department of Physics, Theoretical Physics, 1 Keble Road,\\
Oxford OX1 3NP, England}

\maketitle\abstracts{I briefly discuss recent theoretical advances in the description of  mono-$X$  signals at the LHC.}

\section{Introduction}

The existence of dark matter (DM) provides the strongest evidence  for physics beyond the standard model (SM). DM has been probed using particle colliders, direct detection underground experiments and indirect detection in space telescopes. Despite these intensive searches it has so far proven elusive. We thus must be thorough and creative as we continue the important mission to search for DM. In the coming years, direct and indirect detection experiments will reach new sensitivities, and the LHC will begin operation at 13~TeV after a very successful~(7)~8~TeV run. Taken together these strategies will provide crucial tests of our ideas about the dark sector, and have great potential  to revolutionise  our understanding of the nature of DM. 

\section{Precision mono-jet predictions}

\begin{figure}[!h]
\vspace{5mm}
\begin{center}
\includegraphics[width=0.95\textwidth]{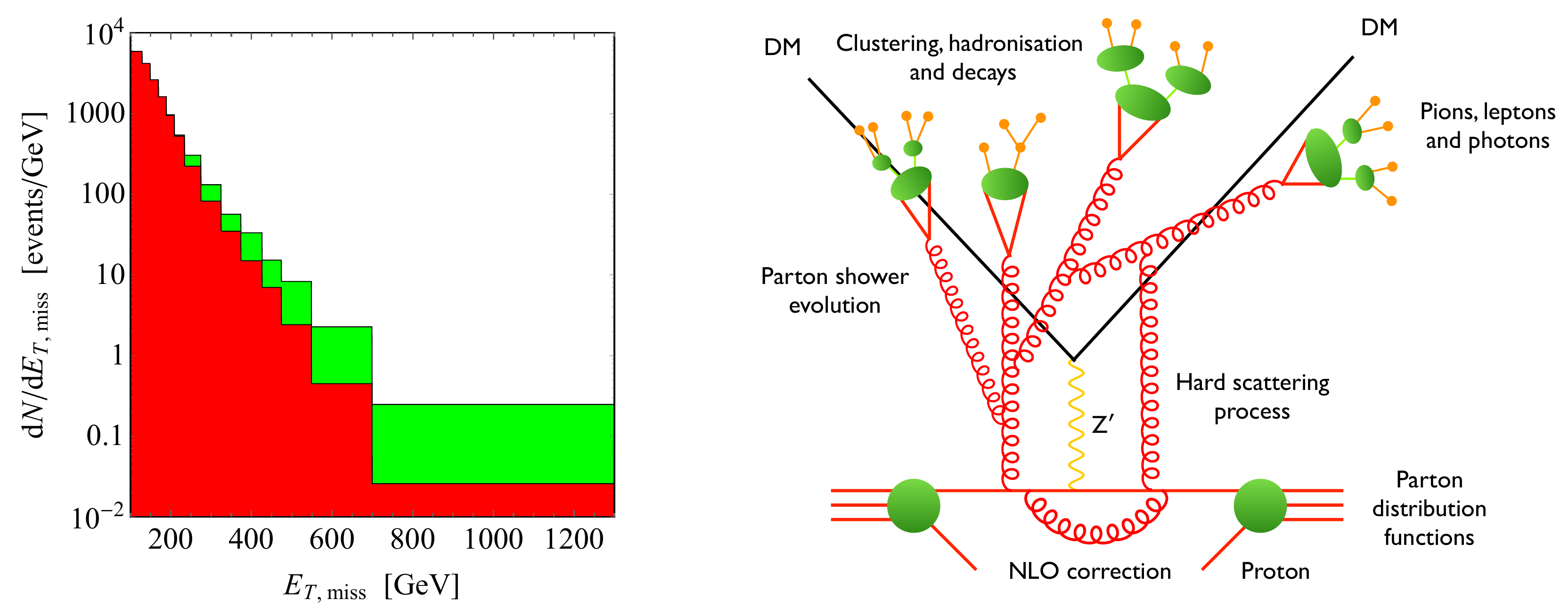}
\caption[]{Left: Comparison between the $\MET$ spectrum of a mono-jet signal arising from DM-pair production~(green) and the leading SM background due to $pp \to j + Z \, (\to \nu \bar \nu)$ (red). Separating the small numbers of signal  events  from the background requires to perform ``tail surgery" at large values of $\MET$. Right: Schematic diagram of all the  ingredients that enter a NLOPS mono-jet prediction.  The propagator labelled $Z^\prime$ indicates a spin-1 resonance that mediates the interactions between the SM quarks and DM.}
\label{fig:1}
\end{center}
\end{figure}

The minimal experimental signature of DM production at the LHC would be an excess of events with a single final-state object  $X$ recoiling against large amounts of missing transverse momentum or energy~($\MET$).  In Run I of the LHC, the ATLAS and CMS collaborations have examined a variety  of such  mono-$X$ signatures involving  jets of hadrons, gauge bosons, top and bottom quarks as well as the Higgs boson in the final state.\cite{Caterina,Deborah} Unfortunately, the SM backgrounds in these searches are large and the $\MET$ spectrum of the signal is essentially featureless although it is slightly harder than that of the SM background. This feature is illustrated on the left  in~Fig.~\ref{fig:1}. Systematical uncertainties were already a limiting factor at Run~I, and will become even more relevant at Run~II. A combination of experimental and theoretical efforts is therefore needed  to improve the reach of future searches. 

From the theoretical side, this requires calculating both background and signal predictions accurately. In the case of mono-jet searches, the dominant backgrounds, resulting from the production of SM vector bosons in association with a jet, have been known to next-to-leading order~(NLO) for a long time.\cite{Giele:1991vf,Baur:1997kz} More recently, attention has been paid to the importance of NLO corrections for the signal process.\cite{Fox:2012ru} These calculations have been implemented into the Monte~Carlo program  {\tt MCFM}.\cite{mcfm} Ideally, the LHC collaborations should be able to use an NLO implementation of the expected DM signal in order to optimise their cuts in such a way that backgrounds are reduced and uncertainties are minimised. For this purpose, a parton-level implementation is insufficient, because a full event simulation including showering and hadronisation is required. This can be achieved using a NLOPS method,~i.e.~an approach that allows to match consistently a NLO computation with a parton shower~(PS). Utilising the {\tt POWHEG} method,\hspace{0.5mm}\cite{Nason:2004rx,Frixione:2007vw} a NLOPS calculation, which permits the automatic generation of mono-jet events in  spin-0 and spin-1 simplified models with $s$-channel exchange, has been performed and is now publicly available.\cite{Haisch:2013ata} A pictorial representation of all the  ingredients needed to achieve NLOPS accuracy  is shown on the right-hand side in Fig.~\ref{fig:1}.

\begin{figure}[!t]
\begin{center}
\includegraphics[width=0.95\textwidth]{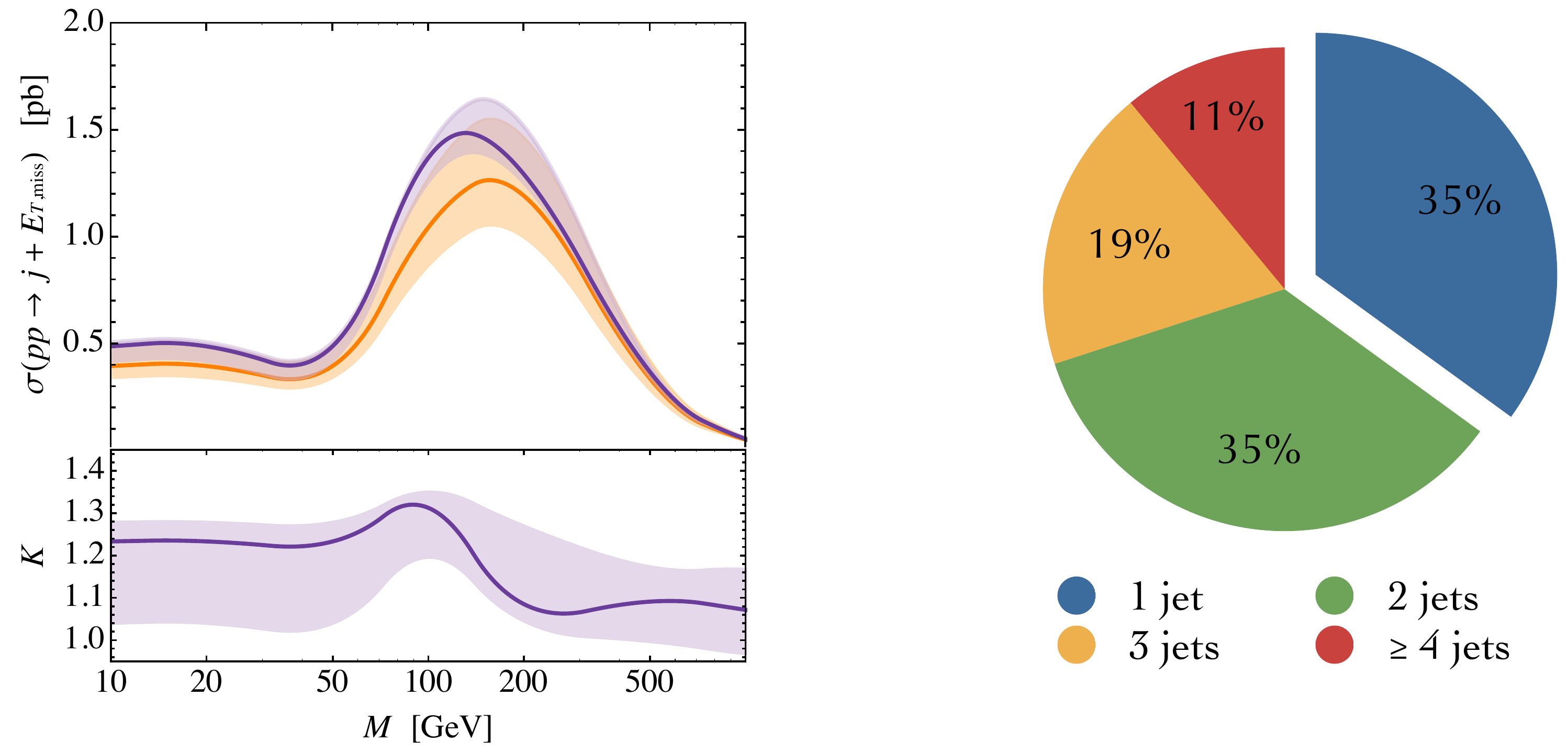}
\caption[]{Left:  NLOPS (purple  curve and band) and LOPS (orange curve and band) mono-jet cross sections as a function of the vector mediator mass $M$ employing CMS cuts.\cite{Khachatryan:2014rra} The resulting $K$ factor,~i.e.~the ratio between the NLOPS and LOPS cross section, is also given. The shown predictions assume a Dirac DM mass $m_\chi = 50 \, {\rm GeV}$ and a total decay width $\Gamma_V = M/3$ of the mediator. Right: Fractions of events with exactly $1$, $2$, $3$ and $4$ or more jets relative  to the total jets + $\MET$ cross section. All numbers correspond to vector interactions, CMS cuts and $m_\chi = 10 \, {\rm GeV}$.}
\label{fig:2}
\end{center}
\end{figure}

For the cuts used by  ATLAS\hspace{0.5mm}\cite{Aad:2015zva} and CMS\hspace{0.5mm}\cite{Khachatryan:2014rra} in their latest analyses, one finds that the mono-jet cross sections at NLOPS are always very similar to the leading order plus PS (LOPS) cross sections,\cite{Haisch:2013ata} which have been used in Run~I to set bounds on the couplings of DM to quarks and gluons. In the case of a vector mediator this feature is illustrated in the left panel of Fig.~\ref{fig:2}. To understand the smallness of higher-order QCD effects, we show on the right-hand side of the same figure the fraction of events of the total jets + $\MET$  cross section with exactly $1$, $2$, $3$ and 4 or more jets, employing CMS cuts.\cite{Khachatryan:2014rra} One observes that~---~ in spite of the name ``mono-jet search''~---~only 35\% of the events contain a single jet, while 65\%  of the cross section is due to events with 2 or more jets. The large importance of secondary jets, which  are not vetoed  in all recent LHC mono-jet analyses, reduces the impact  of the fixed-order NLO corrections to the~$j + \MET$ channel.  In  turn, the resulting NLOPS bounds are not significantly stronger than those obtained at LOPS, but more reliable, since the NLO corrections reduce the factorisation and renormalisation scale uncertainties of the signal prediction.\cite{Fox:2012ru,Haisch:2013ata}

\section{Spin-0 $\bm s$-channel DM models}

\begin{figure}[!t]
\begin{center}
\includegraphics[width=0.95\textwidth]{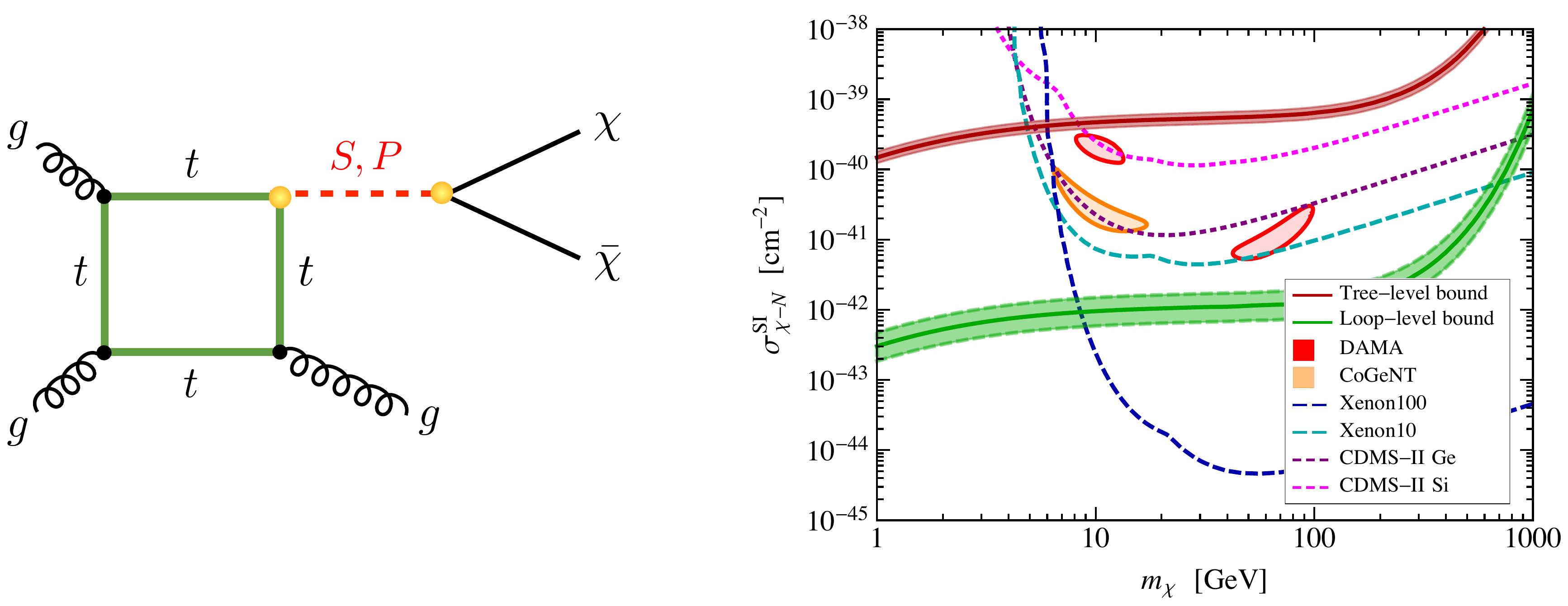}
\caption[]{Left: A typical loop-level diagram with virtual top-quark exchange that leads to mono-jet events. The mediator can be both a scalar ($S$) or pseudoscalar ($P$) resonance. Right: LHC mono-jet bounds on the spin-independent (SI) DM-nucleon cross section $\sigma_{\chi - N}^{\rm SI}$ as a function of the DM mass $m_\chi$ for scalar interactions. While the tree-level mono-jet bound (red curve and band) is too weak to constrain the parameter regions favoured by DAMA and CoGeNT, the loop-level bound (green curve and band) clearly excludes these regions. }
\label{fig:3}
\end{center}
\end{figure}

As we have seen, for spin-1 interactions between DM and the SM, loop corrections do not play an important role, and one may ask if this is a generic feature. The answer is no, because in simplified spin-0 $s$-channel models with Higgs-like couplings, the tree-level $j + \MET$ cross section is small as the heavy-quark luminosities are tiny and the contributions from light quarks are strongly Yukawa suppressed. At the 1-loop level top-quark loops start to contribute to the mono-jet cross section and are expected to lift the Yukawa suppression. A relevant diagram is shown on the right in Fig.~\ref{fig:3}. In fact, including loop contributions in the spin-0 case, increases the  cross section for mono-jet production compared to the tree-level prediction by a factor of around 500 (900) for scalar (pseudoscalar) interactions.\cite{Haisch:2012kf} It is straightforward to translate the constraints arising from mono-jet searches into bounds on the elastic DM-nucleon scattering cross section. For scalar interactions and Dirac DM, the outcome of such an exercise is presented in the left panel of Fig.~\ref{fig:3}. From the figure it is evident that for large DM masses, direct detection experiments give stronger bounds than the mono-jet searches. For $m_\chi \simeq10 \, {\rm GeV}$ however, the constraints become comparable, while below this value the bounds from LHC searches are  superior. One also sees  that the inclusion of 1-loop corrections gives a pertinent improvement of the mono-jet limits, because it excludes the possibility that the DAMA modulation\hspace{0.5mm}\cite{Bernabei:2010mq}  or the CoGeNT excess\hspace{0.5mm}\cite{Aalseth:2011wp} arise from the interactions of a heavy scalar mediator.

\begin{figure}[!t]
\begin{center}
\includegraphics[width=0.95\textwidth]{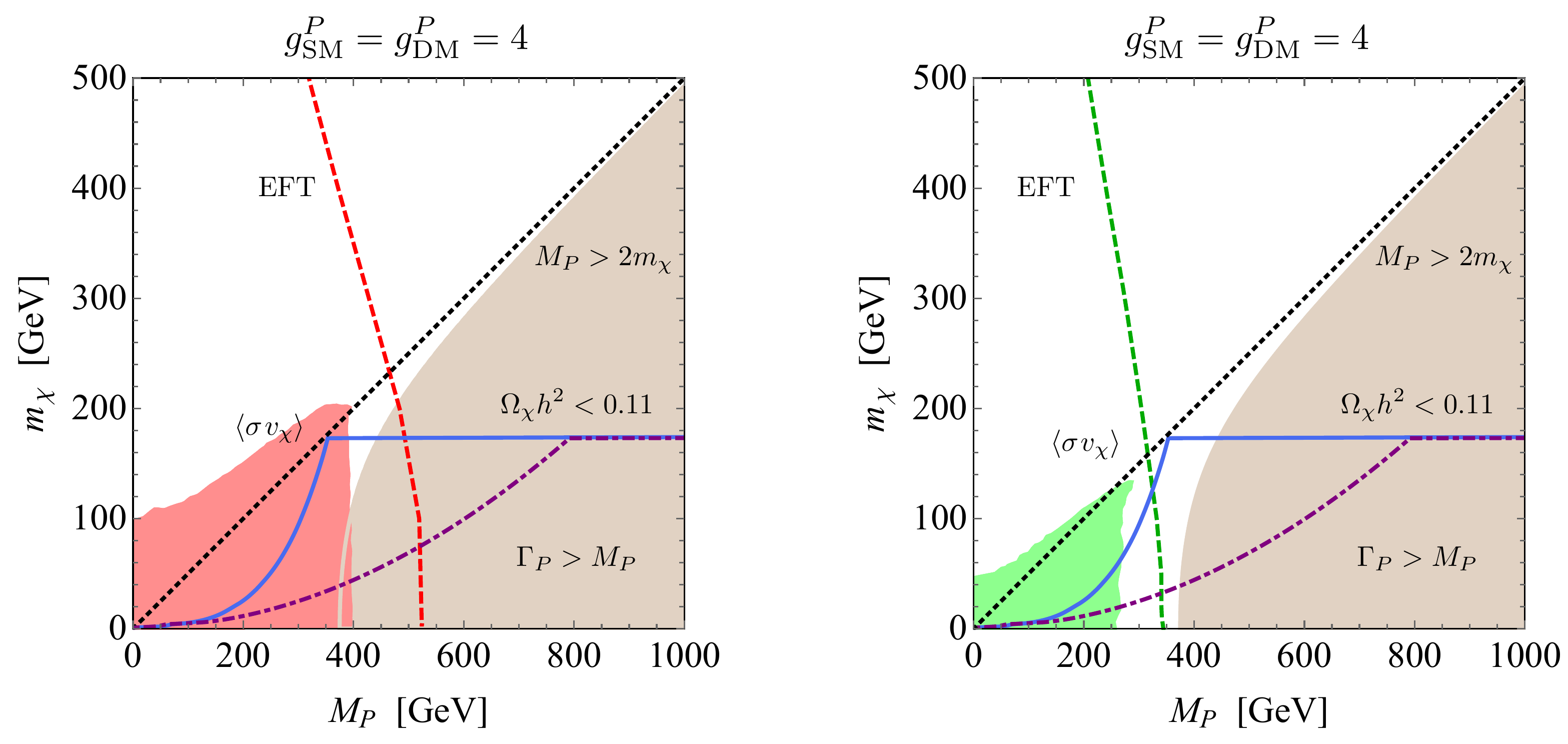}
\caption[]{Exclusion contours at $95\%$~confidence level~(CL)  for pseudoscalar mediators following from Run~I data on $j + \MET$  (red region in left panel) and $t \bar t  \, (\to jbl\nu) + \MET$~(green region in right panel).  The couplings have been fixed to $g_{\rm SM}^P = g_{\rm DM}^P =4$. The regions with $\Gamma_P > M_P$~(brown contours), the parameter spaces with $\Omega_\chi h^2 < 0.11$~(dot-dashed purple curves), the effective field theory (EFT) limits~(dashed red and green curve) and the regions with $M_P > 2 m_\chi$~(dotted black lines) are also shown. The present Fermi-LAT 95\% CL limit\hspace{0.5mm}\cite{Ackermann:2013yva} on the total velocity-averaged DM annihilation cross section $\left \langle  \sigma \hspace{0.25mm} v_\chi \right \rangle$ is indicated by the solid blue curves.}
\label{fig:4}
\end{center}
\end{figure}

A second way to probe spin-0 interactions between DM and top quarks  relies on detecting the top-quark decay products that arise from the tree-level reaction $t \bar t + \MET$.\cite{Deborah,Cheung:2010zf,Lin:2013sca,Artoni:2013zba,Buckley:2014fba,CMS:2014mxa,CMS:2014pvf,Aad:2014vea,Khachatryan:2015nua} Since  the channels $j + \MET$ and $t \bar t + \MET$  test the same interactions, an obvious question to ask  is, which search  sets stronger constraints after Run~I, and to compare their  reach at future stages of LHC operation. By scanning the full 4-dimensional parameter space of the simplified spin-0 models, it has been shown that for both the scalar and the pseudoscalar case the current ATLAS and CMS searches cannot  exclude parameters arising from purely weakly-coupled theories.\cite{Haisch:2015ioa} The scan in addition revealed  that the $j + \MET$ searches in general exclude more parameter space than the $t \bar t + \MET$ searches.  These features are illustrated in the two panels of Fig.~\ref{fig:4}. At the $14 \, {\rm TeV}$  LHC, one finds that  the shapes of the exclusion contours remain qualitatively the same, but that the bounds that one  should be able to set  will improve notable compared to the limits obtained at $8 \, {\rm TeV}$. Still in the initial stages of data taking only model realisations in which the mediators have masses not too far above the weak scale~($M_{S,P} < 1 \, {\rm TeV}$) and couple strong enough to the SM ($g_{\rm SM}^{S,P} > 1$) can be explored.\cite{Haisch:2015ioa}   Since for realistic cuts  the fiducial $pp \to\bar t t \, (\to jbl\nu) + \MET$ cross section is much smaller than that of $pp \to j + \MET$, the $t \bar t + \MET$ channel will only become competitive to the mono-jet signature at the phase-1 and phase-2 LHC upgrades. Realising that the  existing $t \bar t + \MET$ analyses are  all recasts of top-squark searches, the LHC reach might however be improved further by trying to optimise these searches to the specific topology of the $t \bar t + \MET$ signature arising in simplified scalar and pseudoscalar models. 

\section{Angular correlations in DM production}

\begin{figure}[!t]
\begin{center}
\includegraphics[width=0.95\textwidth]{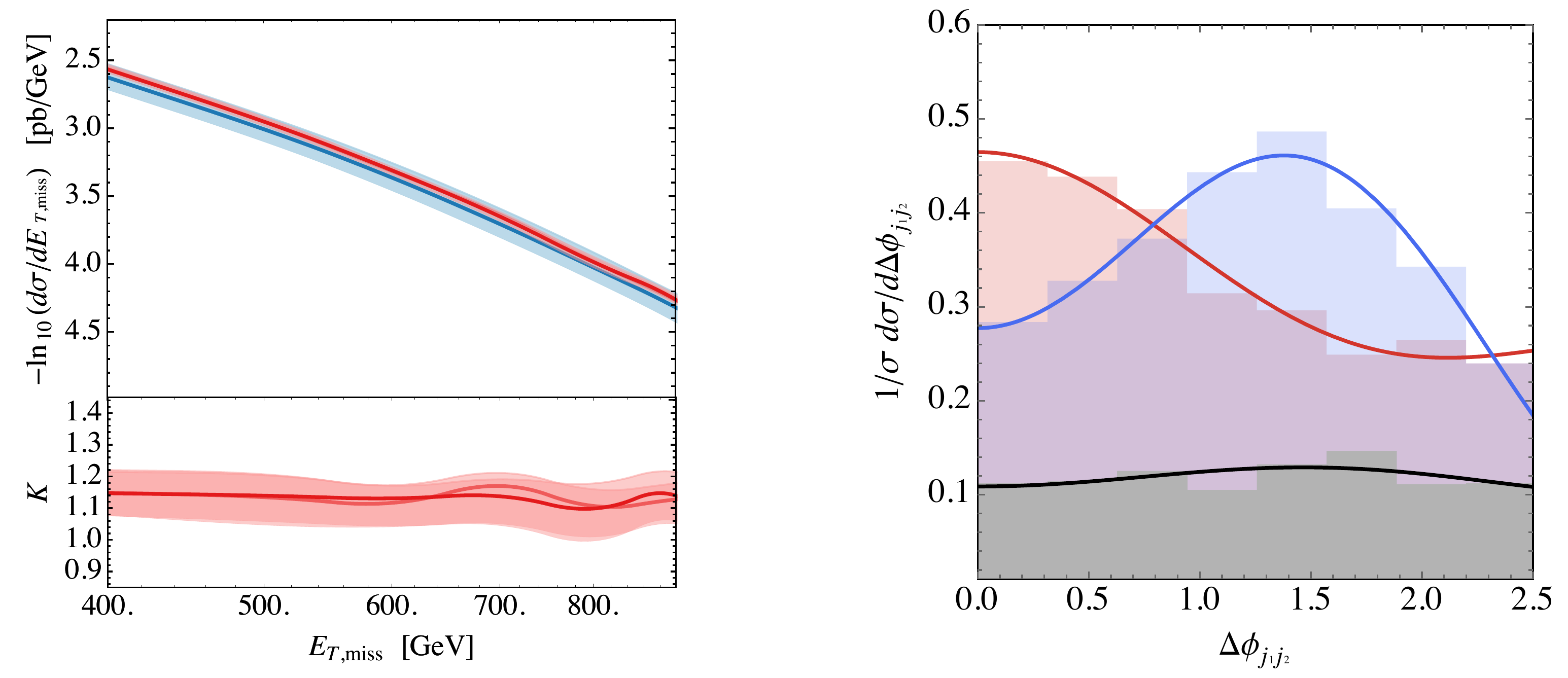}
\caption[]{Left: Comparison of the $\MET$ spectra of a mono-jet signal resulting from vector and axialvector interactions. Overlaid are the LO (blue curves and bands) and NLO (red curves and bands) predictions and the~$K$~factor is also shown. Taking into account scale variations the results are clearly indistinguishable. Right: Normalised~$\Delta \phi_{j_1 j_2}$ distributions for $300 \, {\rm fb}^{-1}$ of $14 \, {\rm TeV}$ LHC data, assuming $m_\chi = 100 \, {\rm GeV}$. The red (blue) histogram shows the signal plus background prediction for the operator $ \bar \chi \chi \hspace{0.5mm}  W_{\mu \nu}^i W^{i, \mu \nu}$ ($\bar \chi \chi \hspace{0.5mm}  W_{\mu \nu}^i \tilde W^{i, \mu \nu}$). The grey bar chart represents the expected SM background, which for better visibility, has been  rescaled by a factor of $1/3$. The solid curves indicate the  best fits of the form $a_0 + a_1 \cos \Delta \phi_{j_1 j_2} +  a_2 \cos \left (2 \hspace{0.25mm} \Delta \phi_{j_1 j_2} \right)$. }
\label{fig:5}
\end{center}
\end{figure}

While the existing $\MET$ searches are well suited to discover DM, they are unlike to provide enough information to determine further DM properties. For instance,  with the existing cut-and-count $j + \MET$  searches it is impossible to distinguish a $\MET$ signal associated to spin-1 vector~($V$) mediator production from one where a axialvector ($A$) resonance furnishes the DM-SM portal by comparing the $\MET$ spectrum of the two different interactions. In fact, from the left panel of Fig.~\ref{fig:5}, one sees that within  theoretical uncertainties the predictions for $pp \to  j + V \, (\to \chi \bar \chi)$ and $pp \to  j + A \, (\to \chi \bar \chi)$ cannot be told apart. 

Some of these limitations can however be overcome by studying two-particle (or multi-particle) correlations in processes involving $\MET$.\cite{Cotta:2012nj,Haisch:2013fla,Crivellin:2015wva} In the case of a mono-jet signal, a sensitive probe of the Lorentz structure of the DM-SM interactions is provided by the jet-jet azimuthal angle difference $\Delta \phi_{j_1 j_2}$ in $2j + \MET$ events.\cite{Haisch:2013fla} The same observable can be used to test the structure of couplings between  pairs of DM particles and gauge bosons.\cite{Cotta:2012nj,Crivellin:2015wva} To demonstrate the power of angular correlations in characterising  the portal  couplings, the normalised $\Delta \phi_{j_1 j_2}$ spectra for two dimension-7  operators with different CP properties is depicted on the right-hand side in Fig.~\ref{fig:5}. The sine-like (cosine-like) behaviour of the modulation in the azimuthal angle distribution corresponding to $ \bar \chi \chi \hspace{0.5mm}  W_{\mu \nu}^i W^{i, \mu \nu}$ ($ \bar \chi \chi \hspace{0.5mm}  W_{\mu \nu}^i \tilde W^{i, \mu \nu}$)~---~with $ W_{\mu \nu}^i $ the $SU(2)_L$ field strength tensor and $\tilde W^i_{ \mu \nu}$ its dual~---~is clearly visible in the figure. The SM background is instead close to flat in the angle $\Delta \phi_{j_1 j_2}$. Last but not least, in the case of $t \bar t + \MET$ production  the pseudorapidity difference  $\Delta \eta_{b_1 b_2}$ ($\Delta \eta_{l^+ l^-}$) between the two bottom-quark jets~(charged leptons) that result from the top-quark decays may be used to disentangle scalar from pseudoscalar interactions.\cite{UH} These three examples show that studies of the correlations of the SM final state particles in $\MET$ events offer unique opportunities to probe the DM-SM interactions, making any dedicated effort at  LHC Run~II in this direction more than welcome. 

\section{Conclusions}

With the start of LHC Run~II, collider searches for $\MET$ signatures are soon to explore new territory, and the large statistics expected at the phase-1 and phase-2 upgrades at $14 \, {\rm TeV}$  have the potential to radically change our understanding of  DM. New theoretical developments that allow for a better description of both signals and backgrounds  have to go along with the experimental advances  in order to exploit the full physics potential of the LHC. Harnessing the ideas  discussed here may play a key  role  in this effort. 

\section*{Acknowledgments}

I thank Andreas~Crivellin, Anthony~Hibbs, Felix~Kahlhoefer, Emanuele~Re and James Unwin for collaborations on various aspects of DM physics. I am furthermore grateful to the organisers of Moriond Electroweak 2015 for the invitation to a great and very enjoyable conference. The warm hospitality and support of the CERN theory division is also acknowledged.

\end{document}